# Improved Stability Design of Interconnected Distributed Generation Resources

Mahmood Saadeh and Roy McCann
Dept. of Electrical Engineering
University of Arkansas
Fayetteville, AR, USA

*Abstract*—**This work provides a design method for achieving a specified level of stability for inverter-based interconnected distributed generation. The stability of parallel connected distributed energy resources determined from a linearized state-space model of the inverter dynamics that includes the admittance matrix of the interconnecting distribution lines. Each inverter uses a localized droop control scheme with the associated voltage and frequency measurements obtained through the application of an enhanced phase locked loop. Previous work on this topic has focused on single inverters connected to an infinite bus without modeling of delays from a phase locked loop implementation. This proposed method overcomes both of these limitations of previous research. A detailed large-signal simulation of a three-bus interconnected power system is analyzed under two different network admittance values. Results confirm the effectiveness of the proposed stability design method.**

## I. INTRODUCTION

As increasing levels of renewable energy sources such as wind and solar come on-line, there is a corresponding increase in concerns related to the design and stability of interconnected inverter systems. Computational stability method have been proposed as in [1] and [2]. There are a number of complex issues related to the coordination of inverter based generation as mentioned in [3] and [4]. Leveraging the extensive experience with conventional synchronous generators, mimicking this behavior has been widely adopted [5]. However, the modeling of power electronic inverters differs from synchronous generators, for example as in [6]. This research effort builds upon that presented by Coelho et al in [7] for developing linearized models of interconnected inverter systems which includes line impedances as part of the dynamics. This provides a means to directly account for the inverter and distribution grid interactions as part of the stability design. The contribution of this paper is that the system is extended to a general three-bus network, whereas [7] only addressed a two-bus system. In addition, this paper presents the use of the enhanced phase-locked loop developed by Karimi-Ghartemani et al in [8]-[9] which allows for robust performance in measuring electrical frequency during power system transients. The overall method of an extended three-bus system with the E-PLL is assessed with a detailed nonlinear simulation that confirms the effectiveness of the proposed stability design method. In the detailed nonlinear simulation the distribution lines were modeled using a pi section line, but are modeled as inductive impedances for the system modeling, and the stability design for simplicity as distribution lines are typically short lines, and the capacitive elements of the pi section line are dominated by the large load impedances $Y_1$, $Y_2$, and $Y_3$.

## II. SYSTEM MODELING

### A. Linearization of a Single Inverter

The approach taken in this research is to linearize the inverter dynamics and assess the eigenvalue for the closed-loop system. This provides a method of designing the droop-controller coefficients. Droop-control is based on the principle that for a single generator the real output power is related to its electrical frequency (and therefore torque angle in steady-state once an equilibrium level has been achieved). It is also noted that reactive power is primarily related to the output terminal voltage magnitude. To summarize, a droop control scheme for a generator is given by

$$\omega = \omega_0 - k_p P \, , \tag{1}$$

$$E = E_0 - k_v Q . \tag{2}$$

The coefficients $k_p$ and $k_v$ are design constants that determine the stability of the overall interconnected inverter system. For inverters, the power electronics can synthesize nearly instantaneously any prescribed voltage and frequency. In order to mimic the behavior of a turbine generator, the inverter first measures the system frequency and makes adjustments with respect to a nominal value (e.g., variations from 60 Hz). This can be cast as the inverter real and reactive power as occurring in response to a measured value that has a first order transfer function characteristic,

$$P_{meas}(s) = \frac{\omega_f}{s+\omega_f} P(s) \, , \tag{3}$$

$$Q_{meas}(s) = \frac{\omega_f}{s+\omega_f} Q(s) \, . \tag{4}$$

And the measurement typically is implemented as a phase-locked loop (PLL). The corresponding linearized dynamics are as given in [7],

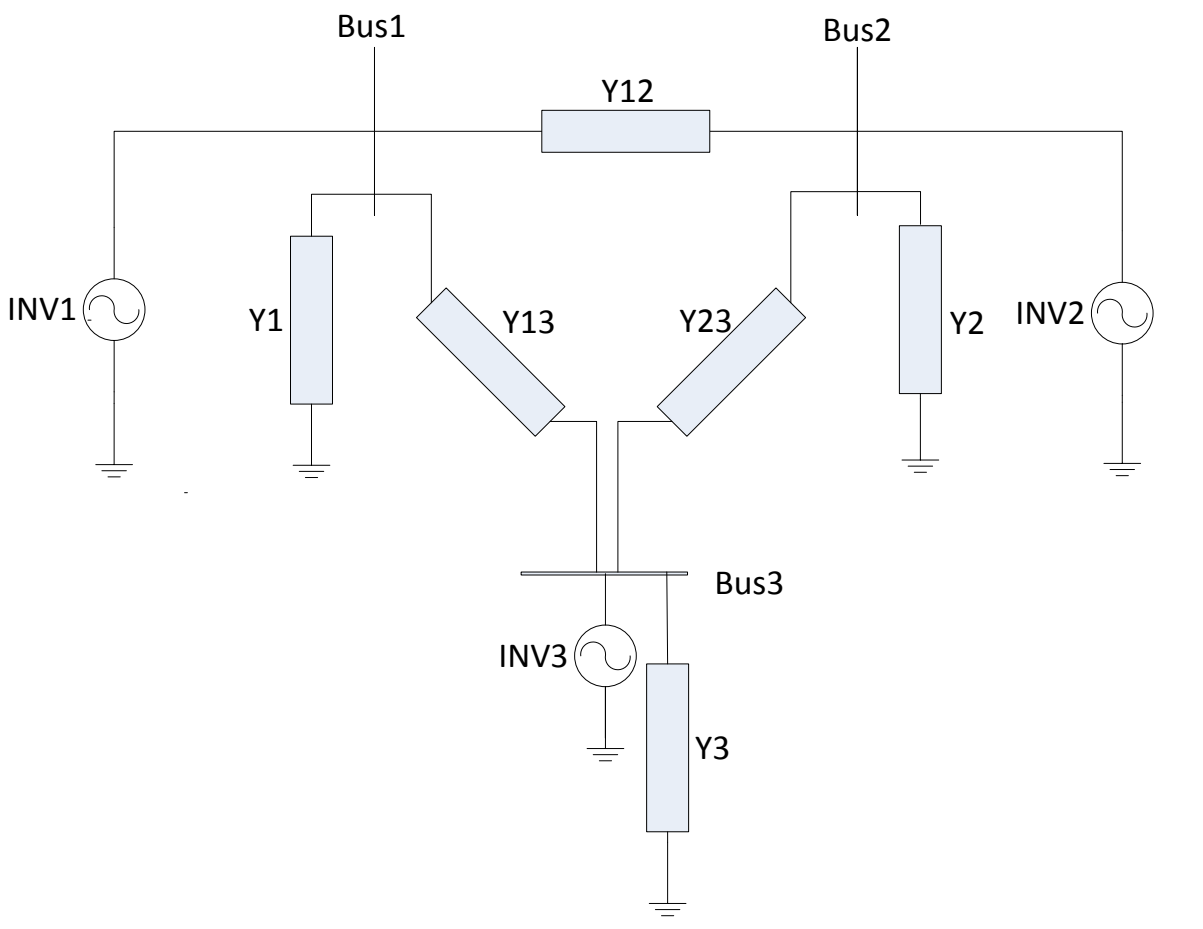

Figure 1. Three-bus inverter network.

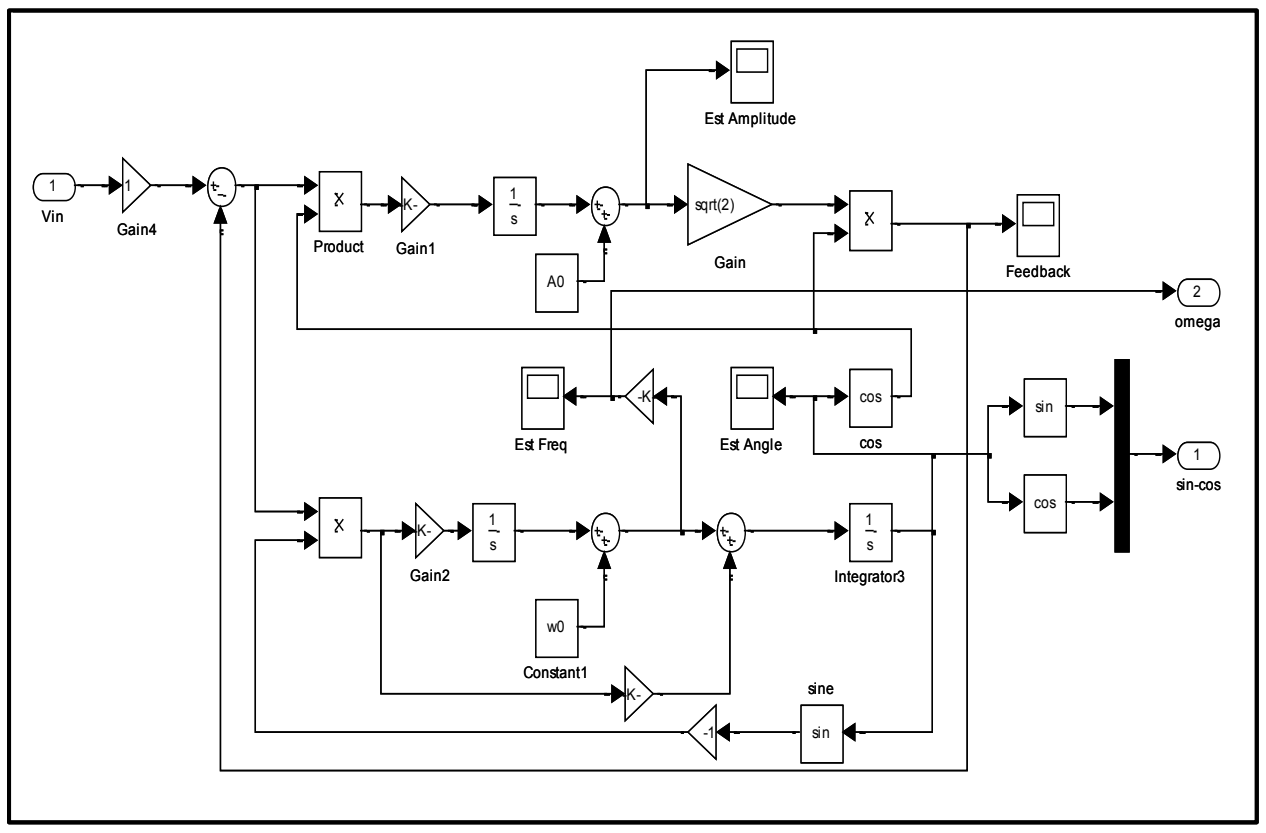

Figure 2. Enhanced phased-lock loop.

$$\Delta\omega(s) = -\frac{k_p\omega_f}{s+\omega_f}\Delta P(s), \tag{5}$$

$$\Delta E(s) = -\frac{k_v\omega_f}{s+\omega_f}\Delta Q(s). \tag{6}$$

The time-domain expressions for (5) and (6) are,

$$\Delta\dot{\omega} = -\omega_f\Delta\omega - k_p\omega_f\Delta P, \tag{7}$$

$$\Delta\dot{E} = -\omega_f\Delta E - k_v\omega_f\Delta Q, \tag{8}$$

in which Δ denotes incremental changes (linearized) of the variable from the equilibrium point. Using the same methods and nomenclature as given by Coelho in [7], it can be shown that the linearized dynamics of each inverter with a first order approximation of the PLL measurement can be expressed in state-space form as

$$\begin{bmatrix}\Delta\dot{\omega}_i\\ \Delta\dot{e}_{di}\\ \Delta\dot{e}_{qi}\end{bmatrix} = [M_i]\begin{bmatrix}\Delta\omega_i\\ \Delta e_{di}\\ \Delta e_{qi}\end{bmatrix} + [C_i]\begin{bmatrix}\Delta P_i\\ \Delta Q_i\end{bmatrix}. \tag{9}$$

.

### *B. Three-Bus Inverter Power System*

The work of Coelho et al in [7] only considered two area systems, of which one was assigned a reference angle. This paper extends the results to a three bus system for which each inverter is of similar size. Thus, there is no simplification of a reference bus to provide an overall system reference voltage, frequency and angle. A general three-bus system is analyzed and is shown in Fig. 1, where its nodal admittance matrix is

$$\begin{bmatrix}\vec{I}_1\\ \vec{I}_2\\ \vec{I}_3\end{bmatrix} = \begin{bmatrix}Y_1+Y_{12}+Y_{13} & -Y_{12} & -Y_{13}\\ -Y_{12} & Y_2+Y_{12}+Y_{23} & -Y_{23}\\ -Y_{13} & -Y_{23} & Y_3+Y_{13}+Y_{23}\end{bmatrix}\begin{bmatrix}\vec{E}_1\\ \vec{E}_2\\ \vec{E}_3\end{bmatrix} \tag{10}$$

The real and imaginary parts of (10) can be separated,

$$\begin{bmatrix}i_{d1}\\ i_{q1}\\ i_{d2}\\ i_{q2}\\ i_{d3}\\ i_{q3}\end{bmatrix} = \begin{bmatrix}G_{11} & -B_{11} & G_{12} & -B_{12} & G_{13} & -B_{13}\\ B_{11} & G_{11} & B_{12} & G_{12} & B_{13} & G_{13}\\ G_{21} & -B_{21} & G_{22} & -B_{22} & G_{23} & -B_{23}\\ B_{21} & G_{21} & B_{22} & G_{22} & B_{23} & G_{23}\\ G_{31} & -B_{31} & G_{32} & -B_{32} & G_{33} & -B_{33}\\ B_{31} & G_{31} & B_{32} & G_{32} & B_{33} & G_{33}\end{bmatrix} \tag{11}$$

For consistency, the notation used in [7] is used such that this can be compactly written as $[i] = [Y_s][e]$. This can then be linearized with respect to an equilibrium point as

$$[\Delta i] = [Y_s][\Delta e] \tag{12}$$

Breaking out the real and reactive power delivered by each inverter in terms of the conventional direct and quadrature components,

$$P_i = e_{di}i_{di} + e_{qi}i_{qi} \tag{13}$$

$$Q_i = e_{di}i_{di} - e_{qi}i_{qi} \tag{14}$$

The three-inverter circuit is described by

$$\begin{bmatrix}\Delta P_1\\ \Delta Q_1\\ \Delta P_2\\ \Delta Q_2\\ \Delta P_3\\ \Delta Q_3\end{bmatrix} = \begin{bmatrix}i_{d1} & i_{q1} & 0 & 0 & 0 & 0\\ i_{q1} & -i_{d1} & 0 & 0 & 0 & 0\\ 0 & 0 & i_{d2} & i_{q2} & 0 & 0\\ 0 & 0 & i_{q2} & -i_{d2} & 0 & 0\\ 0 & 0 & 0 & 0 & i_{d3} & i_{q3}\\ 0 & 0 & 0 & 0 & i_{q3} & -i_{d3}\end{bmatrix}\begin{bmatrix}\Delta e_{d1}\\ \Delta e_{q1}\\ \Delta e_{d2}\\ \Delta e_{q2}\\ \Delta e_{d3}\\ \Delta e_{q3}\end{bmatrix}$$

$$+\begin{bmatrix}e_{d1} & e_{q1} & 0 & 0 & 0 & 0\\ -e_{q1} & e_{d1} & 0 & 0 & 0 & 0\\ 0 & 0 & e_{d2} & e_{q2} & 0 & 0\\ 0 & 0 & -e_{q2} & e_{d2} & 0 & 0\\ 0 & 0 & 0 & 0 & e_{d3} & e_{q3}\\ 0 & 0 & 0 & 0 & -e_{q3} & e_{d3}\end{bmatrix}\begin{bmatrix}\Delta i_{d1}\\ \Delta i_{q1}\\ \Delta i_{d2}\\ \Delta i_{q2}\\ \Delta i_{d3}\\ \Delta i_{q3}\end{bmatrix} \tag{15}$$

Again following the convention used in [7],

$$[\Delta S] = [I_s][\Delta e] + [E_s][\Delta i]\,. \tag{16}$$

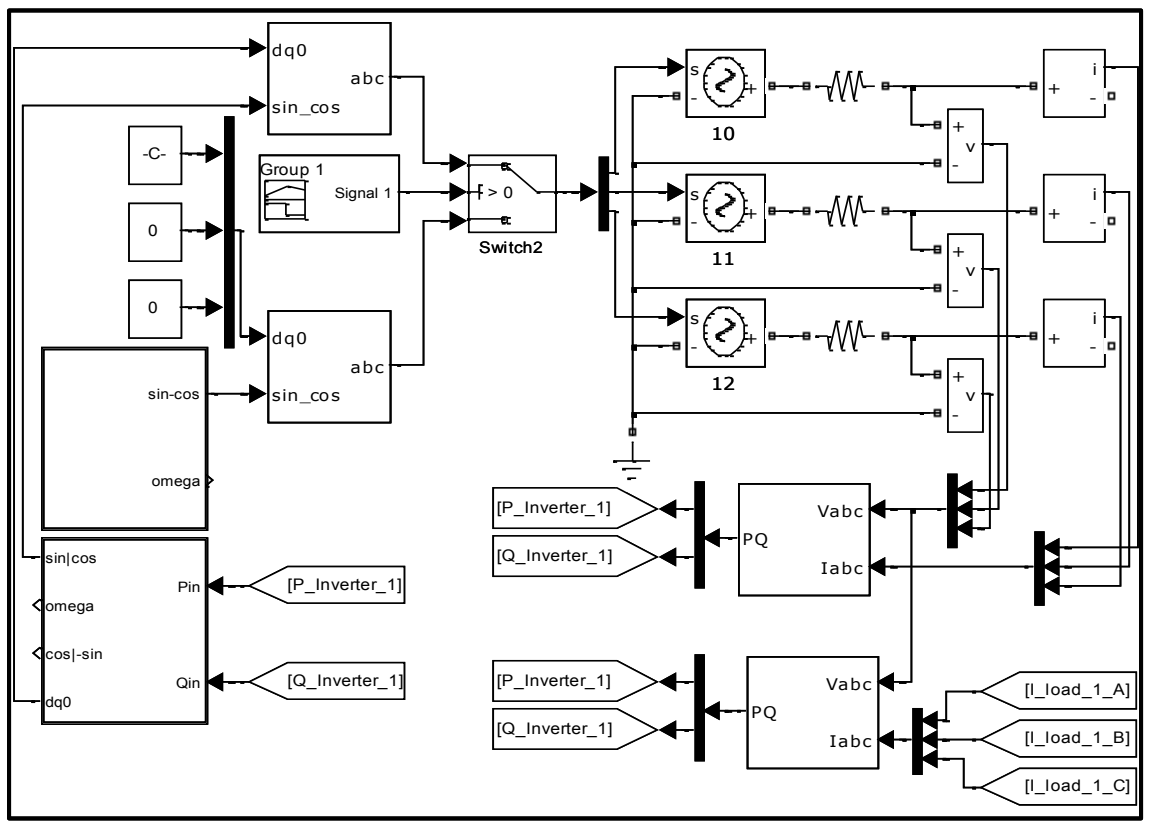

Figure 3. Inverter simulation.

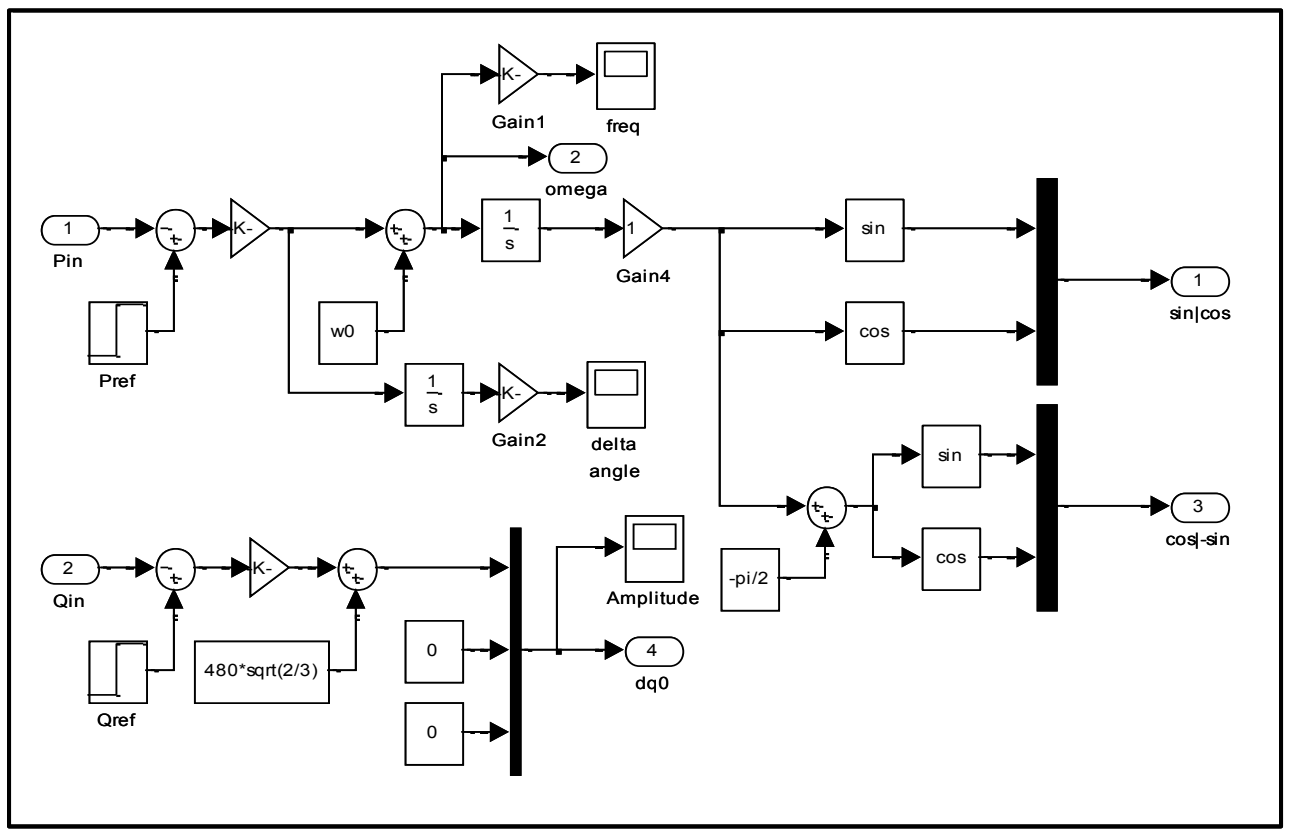

Figure 4. Single area droop control simulation.

The relationship to the admittance matrix is included [7],

$$[\Delta S] = ([I_s] + [E_s][Y_s])[\Delta e] \, , \tag{17}$$

And the overall three-inverter power system dynamics are written as,

$$\begin{bmatrix} \Delta\dot{\omega}_1 \\ \Delta\dot{e}_{d1} \\ \Delta\dot{e}_{q1} \\ \Delta\dot{\omega}_2 \\ \Delta\dot{e}_{d2} \\ \Delta\dot{e}_{q2} \\ \Delta\dot{\omega}_3 \\ \Delta\dot{e}_{d3} \\ \Delta\dot{e}_{q3} \end{bmatrix} = \begin{bmatrix} M_1 & 0 & 0 \\ 0 & M_2 & 0 \\ 0 & 0 & M_3 \end{bmatrix} \begin{bmatrix} \Delta\omega_1 \\ \Delta e_{d1} \\ \Delta e_{q1} \\ \Delta\omega_2 \\ \Delta e_{d2} \\ \Delta e_{q2} \\ \Delta\omega_3 \\ \Delta e_{d3} \\ \Delta e_{q3} \end{bmatrix} + \begin{bmatrix} C_1 & 0 & 0 \\ 0 & C_2 & 0 \\ 0 & 0 & C_3 \end{bmatrix} \begin{bmatrix} \Delta P_1 \\ \Delta Q_1 \\ \Delta P_2 \\ \Delta Q_2 \\ \Delta P_3 \\ \Delta Q_3 \end{bmatrix} \tag{18}$$

and compactly denoted by vector-matrix notation as in [7] by

$$[\Delta\dot{X}] = [M_s][\Delta X] + [C_s][\Delta S]. \tag{19}$$

With substitution, the expression becomes [7]

$$[\Delta\dot{X}] = [M_s][\Delta X] + [C_s]([I_s] + [E_s][Y_s])[\Delta e] \tag{20}$$

The general relationship as in [7] applies for the three-bus system,

$$\begin{bmatrix} \Delta e_{d1} \\ \Delta e_{q1} \\ \Delta e_{d2} \\ \Delta e_{q2} \\ \Delta e_{d3} \\ \Delta e_{q3} \end{bmatrix} = \begin{bmatrix} 0 & 1 & 0 & 0 & 0 & 0 & 0 & 0 & 0 \\ 0 & 0 & 1 & 0 & 0 & 0 & 0 & 0 & 0 \\ 0 & 0 & 0 & 0 & 1 & 0 & 0 & 0 & 0 \\ 0 & 0 & 0 & 0 & 0 & 1 & 0 & 0 & 0 \\ 0 & 0 & 0 & 0 & 0 & 0 & 0 & 1 & 0 \\ 0 & 0 & 0 & 0 & 0 & 0 & 0 & 0 & 1 \end{bmatrix} \begin{bmatrix} \Delta\omega_1 \\ \Delta e_{d1} \\ \Delta e_{q1} \\ \Delta\omega_2 \\ \Delta e_{d2} \\ \Delta e_{q2} \\ \Delta\omega_3 \\ \Delta e_{d3} \\ \Delta e_{q3} \end{bmatrix} . \tag{21}$$

Likewise, the linear relationships are found as along with the state-space dynamics

$$[\Delta e] = [K_s][\Delta X], \tag{22}$$

$$[\Delta\dot{X}] = [A][\Delta X]. \tag{23}$$

In a similar manner to [7] where this is now becomes a three-bus system,

$$[A] = [M_s] + [C_s]([I_s] + [E_s][Y_s])[K_s] \, . \tag{24}$$

This allows a system designer to select the phase-locked loop and the droop-control coefficients to achieve an overall system stability and transient response that is desired for a particular power system deployment. That is, the parameters in (24) contain the two droop coefficients from (1) and (2) and the measurement pole $w_f$ as three design variables, for with well established method such as root-locus techniques, can be used to set the eigenvalues of (24) to a prescribed value for ensuring stability margins of the overall system are achieved.

## III. Enhanced Phased-Lock

The inverter dynamics indicated by (3)-(4) is given by $\omega_f$ which is determined by the response of the phased-locked loop. For this research, the enhanced phased-lock loop (E-PLL) as developed by Karimi-Ghartemani et al in [8] and [9] was used because of its ability to track frequency transients compared to conventional approaches for implementing a PLL. The reader is referred to [8] for the details of the E-PLL operation. To summarize, the following differential equations are solved in real-time for the amplitude, frequency and phase angle [8],

$$\dot{\hat{A}} = 2k_1 e(t) cos\hat{\phi} \, , \tag{25}$$

$$\dot{\hat{\omega}} = -2k_2 sin\hat{\phi} \, , \tag{26}$$

$$\dot{\hat{\phi}} = -2k_3 sin\hat{\phi} + \hat{\omega} \, , \tag{27}$$

$$\text{where } e(t) = x(t) - \sqrt{2}\hat{A}cos\hat{\phi} \; . \tag{28}$$

The implementation of the E-PLL is shown in Fig. 2 and the parameters used in this research are given by,

$A_0 = 480\sqrt{2}$, $k_1 = 200$ , $A_0 k_3 = 2\xi\omega_n$ where $\xi = 0.85$ *and* $\omega_n = 200$, and $A_0 k_2 = (200)^2$.

TABLE I. CASE I PARAMETERS

| Variable | Value | Unit |
|---|---|---|
| Transmission line impedance $Z_{12}$ | *1.5+3i* | *Ω* |
| Transmission line impedance $Z_{13}$ | *0.25+1i* | *Ω* |
| Transmission line impedance $Z_{23}$ | *0.5+4i* | *Ω* |
| Local apparent power load-inverter 1 | *11059+ j*6128* | *VA* |
| Local apparent power load-inverter 2 | *14061+j*6183* | *VA* |
| Local apparent power load-inverter 3 | *7025+j*3462* | *VA* |
| Design time constant for measurement delay Wf | *75.4* | *Rd/s* |
| Frequency droop coefficient kp | *0.0005* | *Rd/s/W* |
| Voltage droop coefficient kv | *0.0005* | *V/VAR* |
| Inverter apparent power 1 | *11073+j*5996* | *VA* |
| Inverter apparent power 2 | *13951+j*5538* | *VA* |
| Inverter apparent power 3 | *7123+j*4296* | *VA* |
| Nominal frequency at operating point | *60* | *Hz* |

TABLE II. CASE I EIGENVALUES

| | |
|---|---|
| $\lambda_1$ | -75.2728 + 0.1148i |
| $\lambda_2$ | -75.2728 - 0.1148i |
| $\lambda_3$ | -69.9089 + 1.3368i |
| $\lambda_4$ | -69.9089 - 1.3368i |
| $\lambda_5$ | -62.8054 |
| $\lambda_6$ | -49.1977 |

## IV. SIMULATION RESULTS

The three bus system previously described was simulated with the enhanced PLL and droop controls active. An inverter system in SimPower Systems [10] is shown in Fig. 3.The implementation of droop-control in Simulink [10] is shown in Fig. 4. The initial start-up from zero initial conditions has a severe transient. The system is run initially for 500 ms to allow the electrical variables of interest to reach an initial steady-state condition. These steady-state values were then used as the initial conditions for running further simulations. The behavior of the system under two separate network conditions was evaluated to explore the droop and E-PLL performance under various conditions.

### A. First Test Case

The first power transient used the parameters and initial conditions as given in Table 1. The associated eigenvalues were computed based on the small-signal analysis and are listed in Table 2. This network was selected in order to have the dynamics of two-sets of complex (i.e., underdamped) eigenvalues. The effects of step changes in power delivered by one inverter are examined to determine how the system responds with droop controls acting through the E-PLL implementations at each inverter control. Two steps were introduced. First, a step of 10% at 0.500 s for the real power supplied by inverter 2, and a step of 10% at 0.500 s for the reactive power supplied by Inverter 3. The resulting change in real and reactive power from each inverter is shown in Fig. 5. The real and reactive power consumed by each load is shown in Fig. 6. The power flow response reaches a new steady state condition within 0.500 ms. The voltage magnitude for Inverter 3 is shown in Fig. 7. This is the peak value of the individual phase to neutral voltage, recalling that this a 480 V line-to-line nominal system. The inverter responds rapidly to measured frequency fluctuations from the E-PLL. The frequency transient at Inverter 2 is shown in Fig. 8, which indicates the expected similarity in the transient response as in Fig. 7. In both the voltage and frequency transients, the new equilibrium values are reached after 80 ms, which corresponds to the designed droop control of 75 ms based on the E-PLL response.

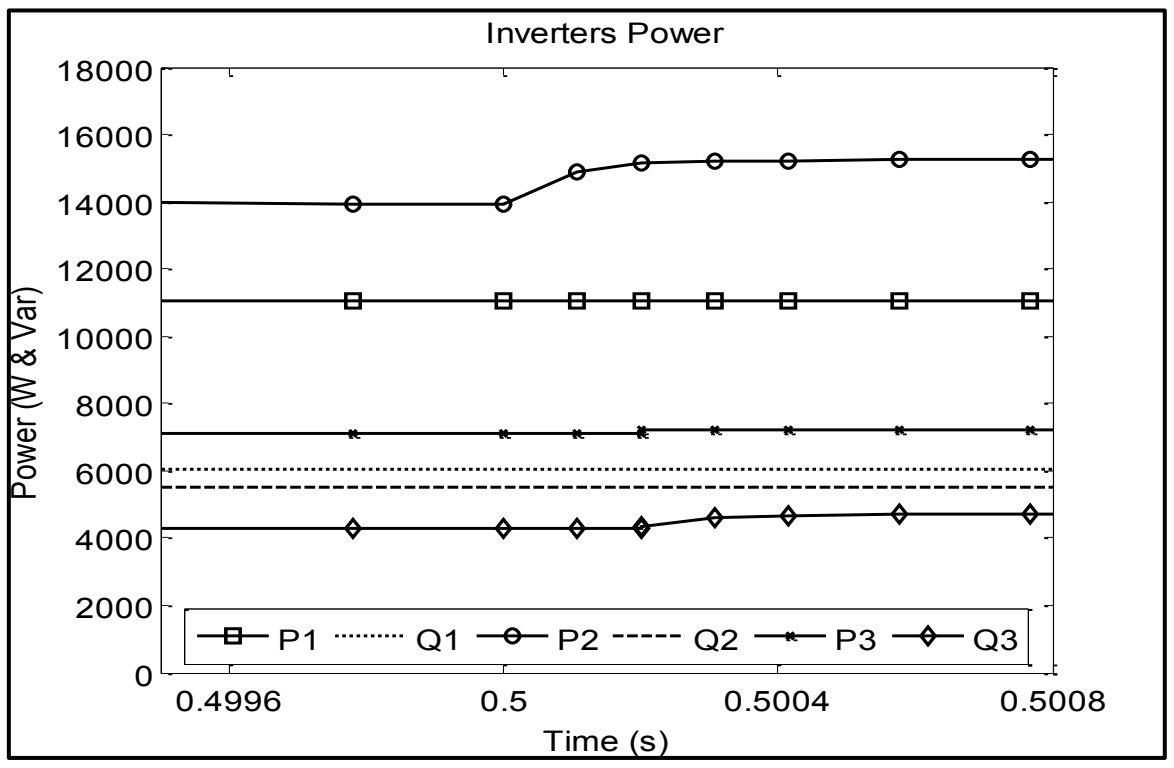

Figure 5. Real and Reactive Power supplied by each inverter.

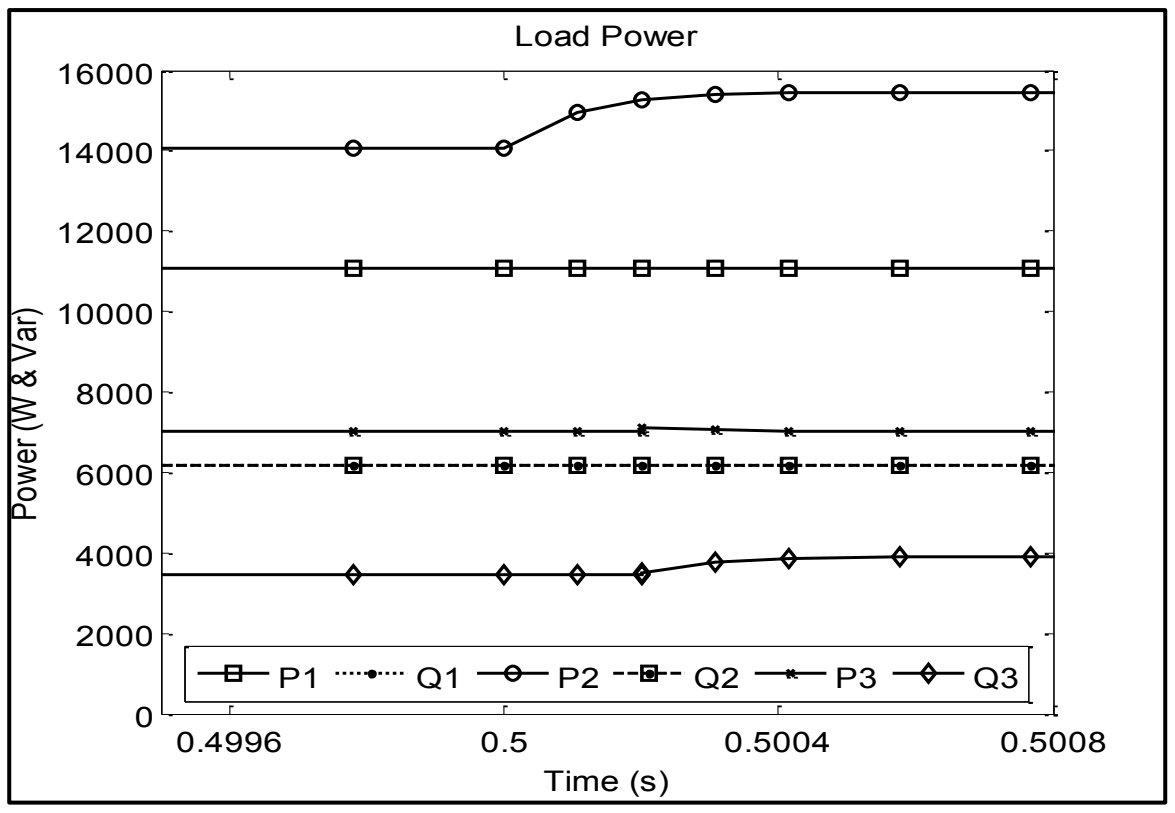

Figure 6. Real and Reactive power consumed by each load.

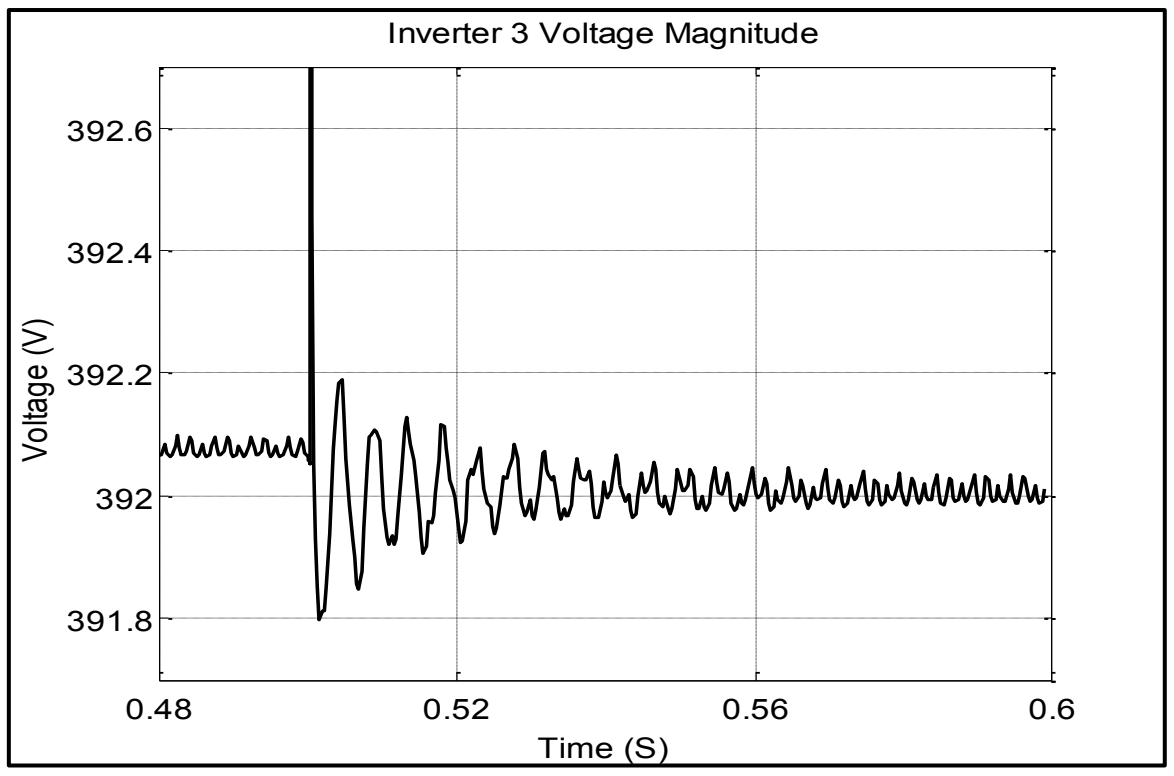

Figure 7. Voltage magnitude of inverter 3.

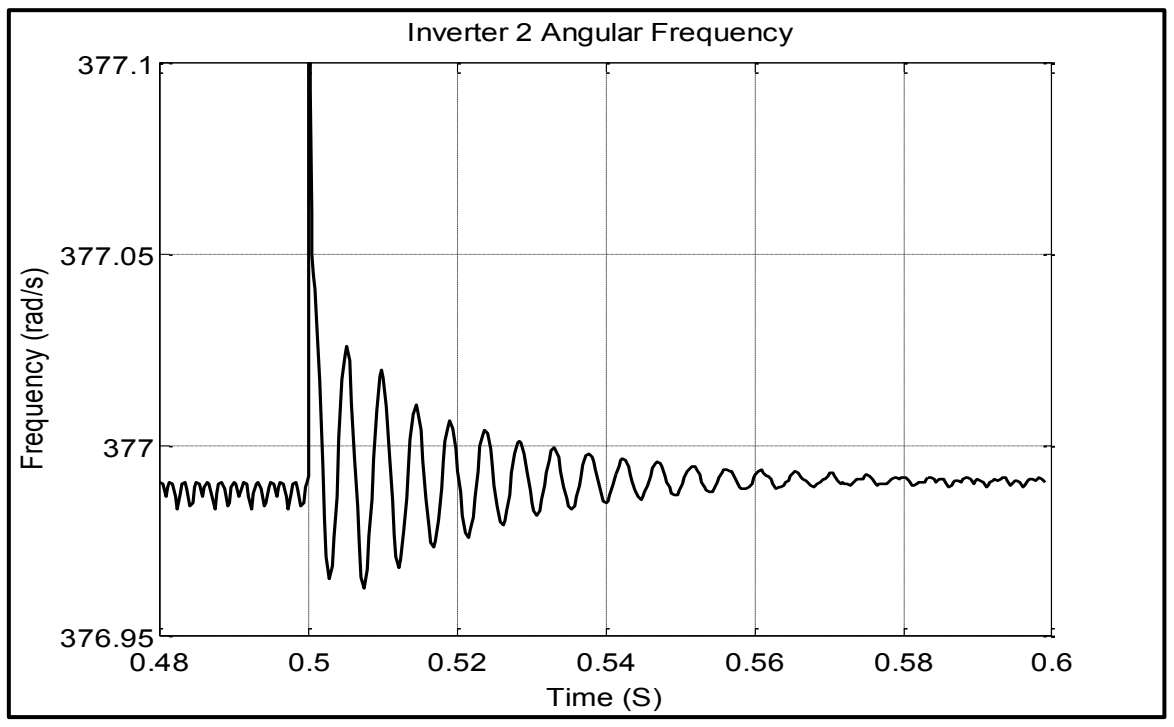

Figure 8. Frequency of inverter 2.

TABLE III. CASE II PARAMETERS

| Variable | Value | Unit |
|---|---|---|
| **Transmission line impedance $Z_{12}$** | *1.5+3i* | *Ω* |
| **Transmission line impedance $Z_{13}$** | *0.25+1i* | *Ω* |
| **Transmission line impedance $Z_{23}$** | *0.5+4i* | *Ω* |
| **Local apparent power load-inverter 1** | *900+ j*400* | *VA* |
| **Local apparent power load-inverter 2** | *750+j*375* | *VA* |
| **Local apparent power load-inverter 3** | *1000+j*450* | *VA* |
| **Design time constant for measurement delay Wf** | *75.4* | *Rd/s* |
| **Frequency droop coefficient kp** | *0.0005* | *Rd/s/W* |
| **Voltage droop coefficient kv** | *0.0005* | *V/VAR* |
| **Inverter apparent power 1** | *900+ j*400* | *VA* |
| **Inverter apparent power 2** | *750+j*375* | *VA* |
| **Inverter apparent power 3** | *1000+j*450* | *VA* |
| **Nominal frequency at operating point** | *60* | *Hz* |

TABLE IV. CASE II EIGENVALUES

| | |
|---|---|
| $\lambda_1$ | -75.7868 |
| $\lambda_2$ | -75.1241 |
| $\lambda_3$ | -70.1289 - 0.8190i |
| $\lambda_4$ | -70.1289 + 0.8190i |
| $\lambda_5$ | -63.1924 |
| $\lambda_6$ | -49.5064 |

### *B. Second Test Case*

Case 2 was simulated in a similar manner to that of case one with the exception that it was simulated with the parameters presented in Table 3, and the real power, reactive power steps were 10%, and 15% respectively. The associated eigenvalues for this network contain on one pair of complex values in Table 4, indicating a system with somewhat greater damping. The same droop and E-PLL values were used as in Case 1. The real and reactive power flows shown in Fig. 9 and Fig. 10 reach a new equilibrium at a similar time as Case 1, although the transient magnitudes are diminished. This is not unexpected given the eigenvalues for the network. The voltage response given in Fig. 11 and the frequency transient in Fig. 12 likewise shown a similar settling time to reach a new equilibrium, however the magnitude of the step-response is somewhat less than Case 1.

## CONCLUSION

This paper derived the three-bus linearized state-space equations for an inverter based power system. This allows for the measurement dynamics associated with a phase-locked loop to be included in the stability analysis of a droop-control based inverter system. The enhanced phase-locked loop was used for deriving the system dynamics, followed by a detailed nonlinear (large signal) simulation. Results under two different networks circuit values demonstrated that the overall system response provides adequate stability when included the overall nonlinear response. Thus, the benefits of the enhanced phase-locked loop with the linearized inverter dynamics were demonstrated.

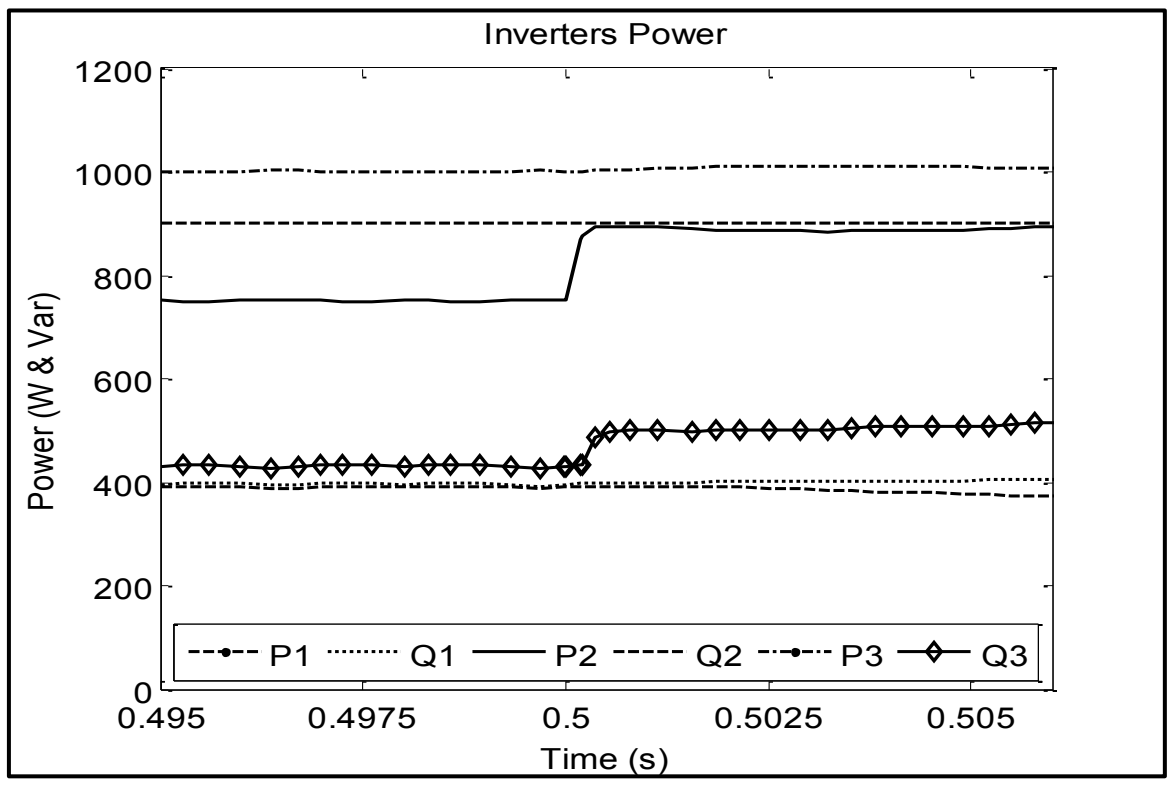

Figure 9. Real and Reactive Power supplied by each inverter.

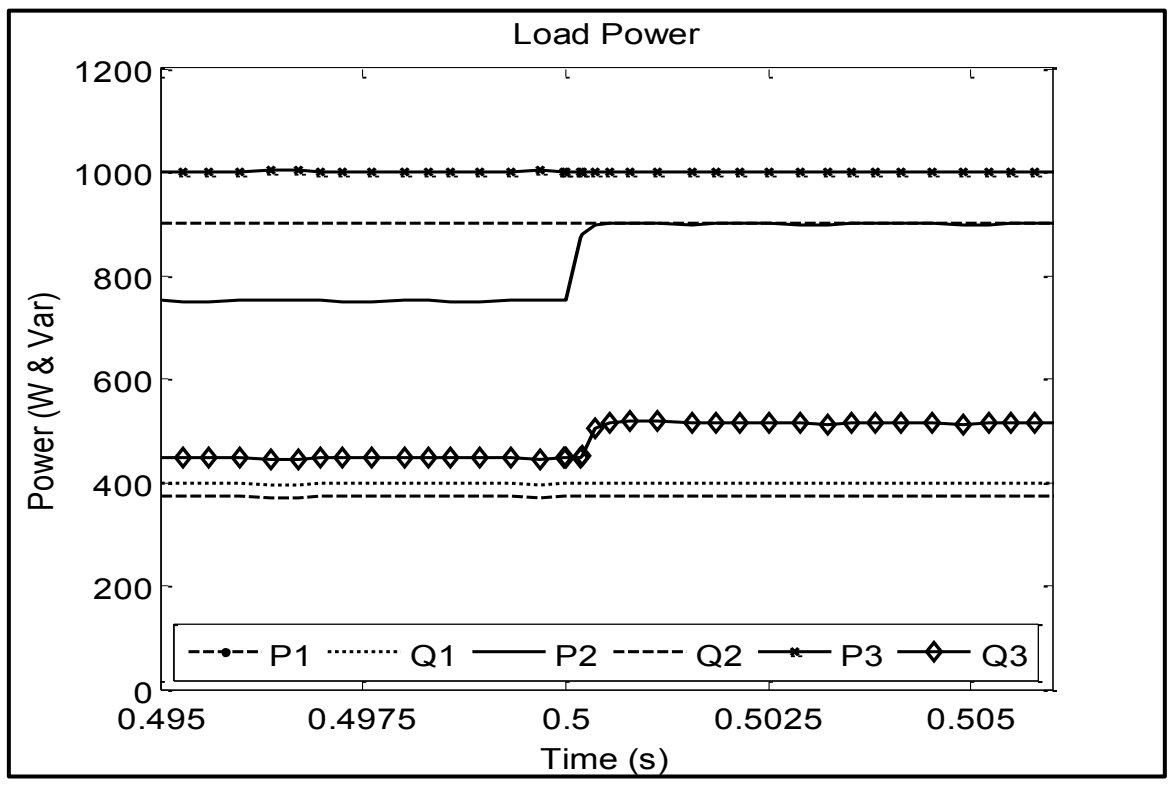

Figure 10. Real and Reactive power consumed by each load.

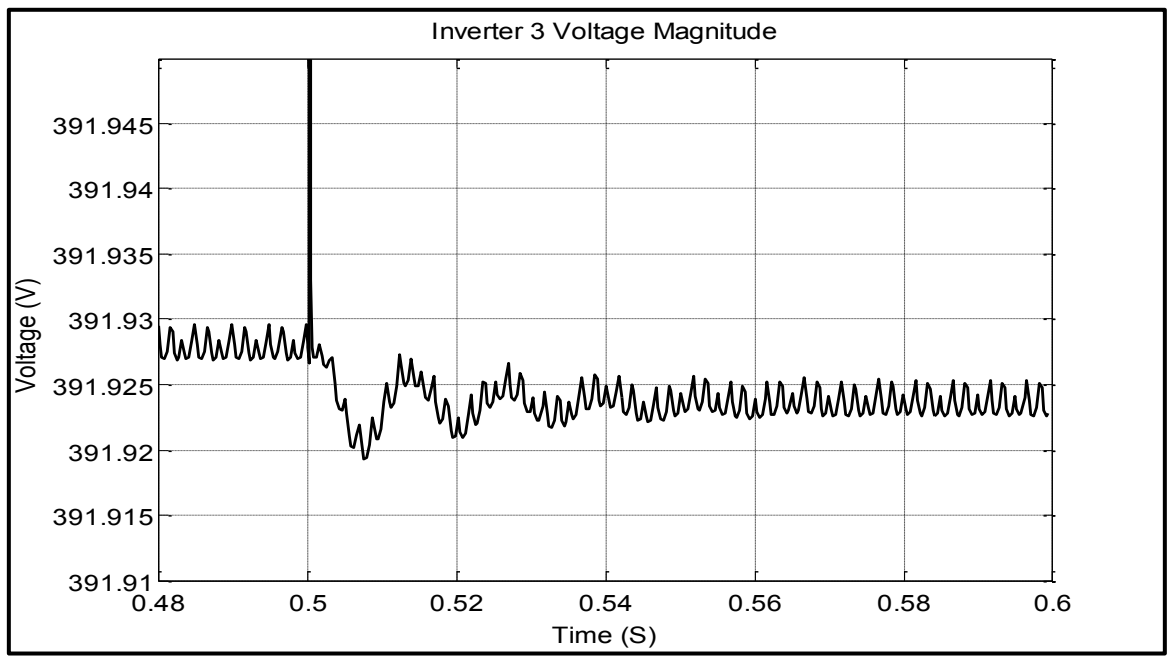

Figure 11. Voltage magnitude of inverter 3.

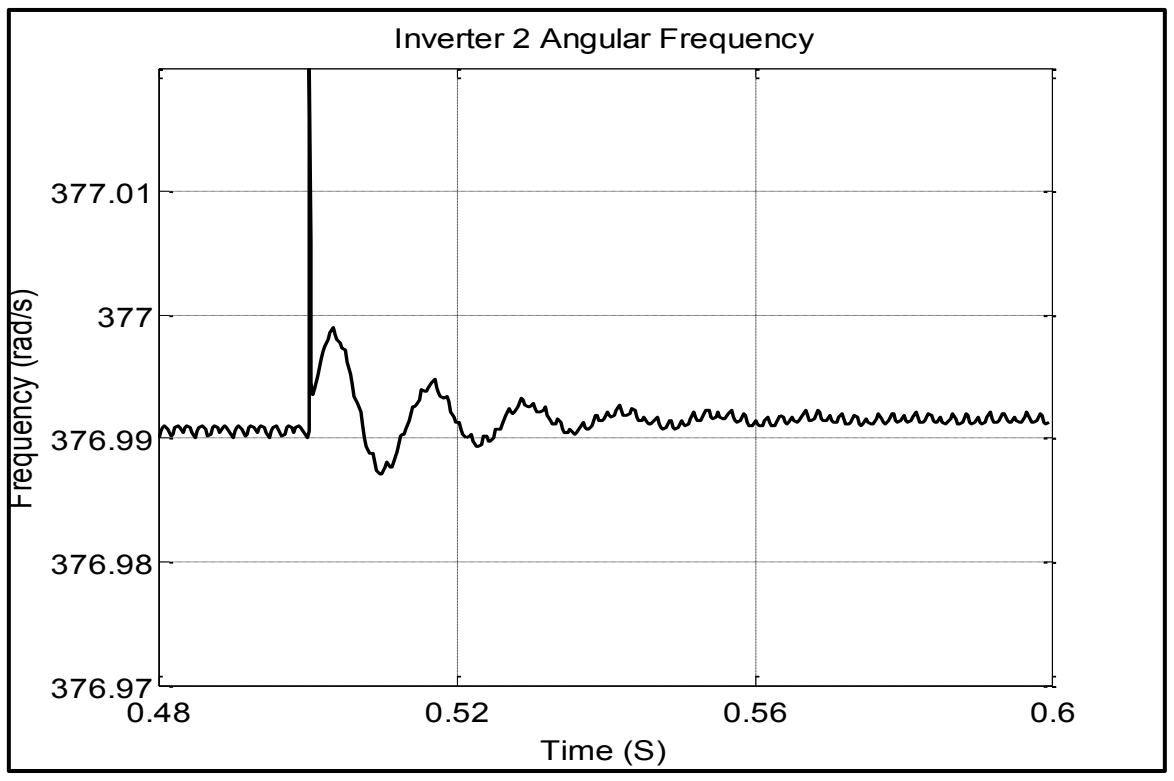

Figure 12. Frequency of inverter 2.

## References

[1] S.V. Iyer, Madhu N. Belur, and Mukul C. Chandorkar, "A Generalized Computational Method to Determine Stability of a Multi-inverter Microgrid," *IEEE Trans. Power Electron.,* vol. 25, no. 9, pp 2420 - 2432 , Sept. 2010.

[2] M. Makhlouf, F. Messai, and H. Benalla, "Modelling and Simulation of Grid-connected Hybrid Photovoltaic/Battery Distributed Generation System," *Canadian Journal on Electrical and Electronics Engineering,* vol. 3, no. 1, Jan 2012.

[3] X. Yu, Z. Jiang, and Yu Zhang, "Control of Parallel Inverter-Interfaced Distributed Energy Resources," *IEEE Energy2030,* 17-18 Nov 2008.

[4] J. Liu, Dragon Obradovic, and A. Monti, "Decentralized LQG Control with Online Set-Point Adaptation for Parallel Power Converter Systems," *IEEE Energy Conversion Congress and Exposition (ECCE),* pp.3174-3179, 12-16 Sept. 2010.

[5] M. Hassanzahraee and A. Bakhshai, "Transient Droop Control Strategy for Parallel Operation of Voltage Source Converters in an Islanded Mode Microgrid," *IEEE 33rd International Telecommunications Energy Conference* (INTELEC), pp.1-9, 9-13 Oct. 2011.

[6] N. Kroutikova, C.A. Hernandez-Aramburo, and T.C. Green, "State-space model of grid-connected inverters under current control mode," *IET Electric Power Applications*, vol.1, no.3, pp.329-338, May 2007.

[7] E. A. Alves Coelho, P. Cabaleiro Cortzio, and P. Donoso Garcia, "Small-Signal Stability for Parallel-Connected Inverters in Stand-Alone AC Supply Systems," *IEEE Trans. Industry Applications*, vol. 38, no. 2, Mar/April. 2002.

[8] M. Karimi-Ghartemani, Boon-Teck Ooi, and Alireza Bakhshai, "Application of Enhanced Phase-Locked Loop System to the Computation of Synchrophasors," *IEEE Trans. Power Delivery*, vol. 26, no. 1, Jan. 2011.

[9] M. Karimi-Ghartemani, M. Mojiri, A. Safaee, J. Walseth, S. Khajehoddin, P. Jain, and A. Bakhshai, "A new phased-locked loop system for three-phase applications," *IEEE Transactions on Power Electronics*, vol. 28, no. 3, pp. 1208-12, March 2013.

[10] The Mathworks, *Matlab & Simulink User's Manual,* Natick MA, January 2013.